\documentclass[prl,aps,showpacs,tightenlines,twocolumn,nofootinbib]{revtex4}
\usepackage{graphicx}

\newcommand{\CL}{{\cal L}}

\newcommand{\CR}{{\cal R}}

\newcommand{\bear}{\begin{array}}  \newcommand{\eear}{\end{array}}
\newcommand{\bea}{\begin{eqnarray}}  \newcommand{\eea}{\end{eqnarray}}
\newcommand{\beq}{\begin{equation}}  \newcommand{\eeq}{\end{equation}}
\newcommand{\bef}{\begin{figure}}  \newcommand{\eef}{\end{figure}}
\newcommand{\bec}{\begin{center}}  \newcommand{\eec}{\end{center}}
\newcommand{\non}{\nonumber}  
\newcommand{\lmk}{\left(}  \newcommand{\rmk}{\right)}
\newcommand{\lkk}{\left[}  \newcommand{\rkk}{\right]}
  
\newcommand{\del}{\partial}  

\newcommand{\bib}{\bibitem} 
\newcommand{\la}{\left\langle} \newcommand{\ra}{\right\rangle}

\def\JL#1#2#3{JETP. Lett. {\bf #1}, #2 (19#3)}

\def\NPB#1#2#3{Nucl. Phys. {\bf B#1}, #2 (19#3)}

\def\PLB#1#2#3{Phys. Lett. B {\bf #1}, #2 (19#3)}
\def\PLBB#1#2#3{Phys. Lett. B {\bf #1}, #2 (20#3)}
\def\PLBold#1#2#3{Phys. Lett. {\bf#1B}, #2 (19#3)}

\def\PRD#1#2#3{Phys. Rev. D {\bf #1}, #2 (19#3)}
\def\PRDD#1#2#3{Phys. Rev. D {\bf #1}, #2 (20#3)}

\def\PRL#1#2#3{Phys. Rev. Lett. {\bf#1}, #2 (19#3)}

\def\PTP#1#2#3{Prog. Theor. Phys. {\bf #1}, #2 (19#3)}

\def\SJNP#1#2#3{Sov. J. Nucl. Phys. {\bf #1}, #2 (19#3)}

\def\ZETFP#1#2#3{Pis'ma Zh. \'Eksp. Teor. Fiz. {\bf #1}, #2 (19#3)}

\newcommand{\lesssim}{ \mathop{}_{\textstyle \sim}^{\textstyle <} }

\begin{document}

\title{Baryogenesis in a flat direction with neither baryon nor lepton
charge}
\author{Takeshi Chiba}
\affiliation{Department of Physics, Kyoto University, Kyoto 606-8502,
Japan}
\author{Fuminobu Takahashi}
\affiliation{Research Center for the Early Universe, University of
Tokyo, Tokyo 113-0033, Japan}
\author{Masahide Yamaguchi}
\affiliation{Physics Department, Brown University, Providence, RI 02912,
USA}
\date{\today}
\begin{abstract}
  We present a new mechanism of spontaneous baryogenesis. Usually such
  mechanisms require a derivative coupling between a scalar field and
  baryon current. In our model, the scalar field instead derivatively
  couples to a current associated with some global symmetry $U(1)_Q$.
  In this case, baryogenesis is still possible provided that an
  interaction exists, which violates the baryon and $U(1)_Q$
  symmetries simultaneously. As a concrete example, we discuss
  baryogenesis in a flat direction with neither baryon nor lepton
  charge.
\end{abstract}

\pacs{98.80.Cq \hspace{4.0cm} KUNS-1834, RESCEU-47/03, BROWN-HET-1351} \maketitle




Baryogenesis is one of the most important challenges in cosmology and
particle physics. Sakharov proposed three conditions to realize
baryogenesis \cite{Sakharov}: (i) an interaction which actually
violates baryon conservation, (ii) violation of $C$ and $CP$
symmetries, (iii) deviation from thermal equilibrium. Thus far, many
scenarios which have been proposed, satisfy the above requirements.
However, a spontaneous baryogenesis mechanism, which was proposed by
Cohen and Kaplan, works even in thermal equilibrium \cite{CK} contrary
to the above conditions. This is because, while Sakharov's conditions
apply for the case where $CPT$ symmetry is conserved, $CPT$ symmetry
is violated in the context of any spontaneous baryogenesis mechanism.
Some applications of spontaneous baryogenesis have recently been
discussed in \cite{DF,LFZ,FNT,Yamaguchi}.

A spontaneous baryogenesis mechanism requires a derivative coupling
between a scalar field $a$ and a baryon current $J_B^\mu$,
\beq
  {\cal L}_{eff} = - \frac{\del_\mu a}{f} J_B^\mu,
\eeq
where $f$ is a cut-off scale. The baryon current $J_B^\mu$ is given
by
\beq
J_B^\mu =
\sum_i B_i j^\mu_i,
\eeq
where $B_i$ and $j^\mu_i$ are, respectively, the baryon number and the
usual number current of the $i$-th field. A nonzero value of $\del_\mu
a$ leads to spontaneous $CPT$ violation since $\del_\mu a$ is odd
under $CPT$ transformations. This enables spontaneous baryogenesis
mechanisms to evade Sakharov's conditions. Assuming that $a$ is
homogeneous, the derivative interaction becomes
\beq
  {\cal L}_{eff} = - \frac{\dot{a}}{f} n_B
                 \equiv \sum_i \mu_i n_i,
\eeq
where the dot denotes differentiation with respect to time, $n_B$ is
the baryon number density, and $n_i$ is the number density of the
$i$-th field. $\mu_i \equiv - \dot{a} B_i/f $ is an effective chemical
potential for the $i$-th field, which creates a bias between baryons
and anti-baryons. Then, according to the effective chemical potential,
the baryon number density is induced while in thermal equilibrium,
\beq
\label{eq:nd}
n_{B}(t) = \sum_i B_i\frac{g_i  \kappa_i  T^3}{6} \left(
\frac{\mu_i}{T} + O\left[\left(\frac{\mu_i}{T}\right)^3\right]
\right)\,,
\eeq
where $g_i$ represents the degree of freedom of the corresponding
fields, $\kappa_i$ is 1 for fermions and 2 for bosons, and the
summation should be taken for light fields in thermal equilibrium. In
fact, baryon asymmetry is generated only when a baryon number
violating interaction actually exists and is in thermal equilibrium.
Thus, the final baryon asymmetry is determined at decoupling of the
baryon number violating interaction.

In the above example, a scalar field derivatively couples to the
baryon current. However, in this Letter, we show that baryogenesis is
still possible even if a scalar field derivatively couples not to
baryon current, but to other currents such as the lepton current
and/or the Peccei-Quinn ($PQ$) current.  Our basic idea is very
simple. Assume that a scalar field derivatively couples to some global
$U(1)_Q$ current which is not necessarily related to baryon current.
Then, if an interaction exists and is in thermal equilibrium, which
violates $U(1)_Q$ and baryon symmetries simultaneously, baryon
asymmetry is generated in addition to the $U(1)_Q$ asymmetry through
the spontaneous mechanism.  This is because the ratio of $U(1)_Q$
charge to baryon charge must follow the ratio of violations of the
interaction. Therefore generation of $U(1)_Q$ asymmetry always
involves generation of baryon asymmetry.

The above idea has a lot of implications. For example, if a scalar
field derivatively couples to lepton current and an interaction
violating both lepton and baryon symmetries is in thermal equilibrium,
baryon asymmetry is generated without resort to sphaleron effects.
More amazingly, even if a scalar field has an interaction with neither
baryon nor lepton charge but only derivatively couples to some global
$U(1)_Q$ current such as $PQ$ current, baryon asymmetry is generated
provided that an interaction which violates both $U(1)_Q$ and baryon
symmetries is in thermal equilibrium.\footnote{In fact, baryogenesis
  is still possible for an interaction which violates both $U(1)_Q$
  and lepton symmetries. In this case, a lepton asymmetry is
  generated, which can be converted into a baryon asymmetry through
  sphaleron effects. As a concrete example, we discuss this
  possibility later in detail.}

To make our ideas clear and concrete, we consider spontaneous
baryogenesis in the context of flat directions with neither baryon nor
lepton charge. The characteristics of flat directions enable
spontaneous baryogenesis to be realized. First of all, due to its
flatness, a flat direction can easily acquire a large vacuum
expectation value, which yields a derivative interaction between its
phase and the symmetry current under which the flat direction is
charged. Second, by virtue of a charge violating term (a so-called
A-term), the phase starts rotating, thus acquiring a nonzero velocity.
Thus, flat directions are suitable for realizing spontaneous
baryogenesis. Use in baryogenesis of these features of flat directions
was first proposed by Affleck and Dine in a slightly different context
\cite{AD}. If a flat direction has baryon and/or lepton number,
rotation of its phase due to an A-term generates baryon and/or lepton
asymmetries as a condensate of the flat direction. After decay of the
flat direction, such asymmetries are transferred to the ordinary
quarks and leptons completing baryo/leptogenesis. However, this
mechanism applies only to flat directions with non-zero $B-L$ charge
because otherwise sphaleron effects wash out the produced baryon
asymmetry. Recently, it was pointed out that this mechanism can be
applied to flat directions with vanishing $B-L$ charge by virtue of
$Q$-balls \cite{Kusenko, Enqvist, Kasuya1}. This is because $Q$-balls
can protect the $B+L$ asymmetry from the sphaleron effects. However,
the mechanism does not work for flat directions with neither baryon
nor lepton charge because no condensation of baryon or lepton number
is produced. In this Letter, we show that baryogenesis is still
possible for such a flat direction by applying the spontaneous
mechanism with our new idea stated above, that is, baryon and/or
lepton asymmetries generated thermodynamically instead of through
condensation. Further implications and details of our ideas will be
discussed in the future publication \cite{TY}.

Now let us go into some details of our scenario. First, we show that
if a flat direction with neither baryon nor lepton number is charged
under another global symmetry and has a nonzero expectation value, its
phase (more precisely, the Nambu-Goldstone (NG) boson associated with
breaking of the global symmetry) derivatively couples to the
corresponding current. One of the famous examples of such a global
symmetry is the $PQ$ symmetry, which is introduced to solve the strong
$CP$ problem of quantum chromodynamics \cite{pq}. For definiteness, we
adopt the supersymmetric DFSZ axion model\footnote{Here we assume that
  the $PQ$ scalar fields, which are responsible for the spontaneous
  $PQ$ symmetry breaking in the present universe, have negative mass
  squared of order of the Hubble parameter squared during inflation.
  Then we can avoid the problem of axion domain walls. In addition,
  cold dark matter (CDM) can also be explained by the axion in our
  scenario as a by-product of adopting the $PQ$ symmetry.}
\cite{DFSZ,Ma}, but it is trivial to extend it to the case with
general global $U(1)$ symmetries.

Though a flat direction is specified by a holomorphic gauge-invariant
polynomial, it is often described by a single complex scalar field
$\Phi \equiv \phi/\sqrt{2}\, e^{i \theta}$. Assuming that a flat
direction has only $PQ$ charge, its phase is given by
\beq
  \theta = \frac{1}{N} \CR \alpha_R,
\eeq
where $N$ is the number of constituent fields, $\CR \equiv \sum_i R_i$
is the sum of the $PQ$ charges, $R_i$ is the $PQ$ charge of each
field, and $\alpha_R$ is the angle conjugate to the generator of the
$PQ$ symmetry. If the flat direction has a nonzero expectation value,
the $PQ$ symmetry is spontaneously broken and a NG boson $a_{R}$
appears, which is given by
\beq
  a_{R} \equiv v_a \alpha_R,
\eeq
where $v_a \sim \la \phi \ra$ is a decay constant. Then, the NG mode
$a_R$ transforms as $a_R \rightarrow a_R + v_{a} \epsilon$ under the
$PQ$ transformation $\alpha_R \rightarrow \alpha_R + \epsilon$.

In order to obtain the interaction between the NG mode $a_R$ and the
other charged fields, we define the $PQ$ current as
\beq
  J^{\mu}_{R} \equiv -\sum_{m'} \frac{\del\CL}{\del(\del_{\mu} \chi_{m'})} 
                      \delta\chi_{m'},
\eeq
where $m'$ denotes all fields with non-zero $PQ$ charges; that is,
$\chi_{m'}$ transforms under $PQ$ symmetry as $\chi_{m'} \rightarrow
\chi_{m'} + \epsilon  \delta\chi_{m'}$ with $\delta\chi_{m'} = i  R_{m'}
\chi_{m'}$. Then, the $PQ$ current is given by
\beq
  J^{\mu}_{R} = v_{a} \del^{\mu} a_R + \sum_{m} R_{m} j^\mu_m,
\eeq
where $m$ denotes all fields with non-zero $PQ$ charges except the NG
mode $a_R$. Current conservation yields the equation of motion for
$a_R$:
\beq
  \del_{\mu} J^{\mu}_{PQ} = v_{a} \del^2 a_R 
                      + \sum_{m} R_{m} \del_{\mu} j^\mu_m = 0.
\eeq
Here the first term in the middle equation is the kinetic term for the
NG mode $a_R$ and the second term can be derived from the following
effective Lagrangian,
\beq
  \CL_{\rm eff} = -\sum_{m} \frac{R_{m}}{v_a}
                   \left( \del_\mu a_R \right) j^\mu_m,
  \label{eq:dercoupling}
\eeq
which yields the derivative interactions between the NG mode $a_R$ and
the other charged fields.

Next, we show that a flat direction has a nonzero expectation value
during and after inflation, and the corresponding NG boson rolls down
along the potential; that is, it acquires a nonzero velocity.
Although there are no classical potentials along flat directions in
the supersymmetric limit, they are lifted by both supersymmetry
breaking effects and non-renormalizable operators. Adopting
gravity-mediated supersymmetry breaking, a flat direction has a soft
mass $m_\phi \sim 1$TeV. Moreover, assuming a non-renormalizable
operator in the superpotential of the form
\begin{equation}
\label{eq:spnr}
    W = \frac{1}{n M^{n-3}} \Phi^n\,,
\end{equation}
the flat direction is further lifted by the potential
\begin{equation}
\label{eq:pnr}
    V_{NR} = \frac{|\Phi|^{2 n-2}}{M^{2 n -6}}\,,
\end{equation}
where $M$ is a cutoff scale. During the inflationary epoch, there is
another contribution to the potential. A flat direction has a negative
mass squared proportional to the Hubble parameter squared, which is
derived from a four-point coupling to the inflaton in the K\"ahler
potential. Then, this negative mass squared term destabilizes the flat
direction at the origin and the flat direction rolls down toward the
minimum of the potential. The minimum of the potential $\phi_{\rm
  min}$ is determined by the balance between the negative mass squared
term and the non-renormalizable potential $V_{NR}$, and is given by
\beq
   \phi_{\rm min} \sim 
     \left(H M^{n-3} \right) ^{\frac{1}{n-2}}.
\label{eq:potential_min}
\eeq

In fact, the above non-renormalizable superpotential not only lifts
the potential but also gives the charge-violating A-terms of the form
\bea
\label{eq:nrA}
    V_{A} &=& a_m \frac{m_{3/2}}{n M^{n -3}} \Phi^n + {\rm h.c.} 
               \non \\
           &=& M_A^4 \cos \lkk k \CR a_R / v_a \rkk,
\eea
where $m_{3/2}$ is the gravitino mass, $a_m$ is a complex constant of
order unity, $M_A$ is the energy scale of the A-term, $k \equiv n/N$,
and we have assumed a vanishing cosmological constant. By virtue of
this A-term, the NG boson $a_R$ begins rotating and acquires a nonzero
velocity given by
\beq
  \left|\dot{a}_R \right| \sim \frac{k {\cal R}}{H v_a} M_A^4,
\eeq
where we used the slow-roll approximation because the inverse
curvature scale of the potential is roughly $\sqrt{m_{3/2} H} \ll H$.
We have also assumed that $a_R$ sits far from the extremum of the
potential by $O(v_a)$.

The final ingredient to realize the spontaneous mechanism is an
interaction which actually breaks the relevant symmetry. In this
Letter, as stated in the introduction, we consider an interaction
which violates multiple symmetries simultaneously, that is, $PQ$ and
lepton symmetries.  Such a violating interaction is characterized by
the amounts of violation of the $PQ$ and lepton symmetries, that is,
$\Delta_R$ and $\Delta_L$. Then, since the produced asymmetries $n_R$
and $n_L$ should be proportional to the amount of each violation, the
simple relation $n_R \Delta_L = n_L \Delta_R$ must be satisfied, or
equivalently,
\beq
\label{eq:constraint}
 \sum_m \Xi_m n_m =0,
\eeq
where we have defined $\Xi_m = R_m \Delta_L-L_m \Delta_R$ and $L_m$ is
the lepton charge of the $m$-th field. However, generally speaking,
the above relation is not satisfied by the usual estimate of the
produced asymmetry given by
\beq
\overline{n}_m (t_D) = \frac{ \kappa_m g_m}{6} \overline{\mu}_m T_D^2
\eeq
with $\overline{\mu}_m \equiv - R_m \dot{a}_R/v_a$. Therefore,
$\overline{\mu}_m$ cannot be interpreted to be a chemical potential of
the $m$-th field. Instead, we consider the projection of
$\{\overline{\mu}_m\}$ onto the parameter plane perpendicular to
$\{\Xi_m\}$ in order to satisfy the above constraint. Such a
projection $\tilde{\mu}_m$ is given by
\beq
\label{eq:mutilde}
\tilde{\mu}_m
\equiv \overline{\mu}_m -
\frac{\left(\overline{\mu} \cdot \Xi \right)}{\Xi^2}
\Xi_m,
\eeq
where we adopt the general shorthand
\bea
Y^2 &\equiv& \sum_m \kappa_m g_m Y_m^2,\non\\
Y \cdot Z &\equiv& \sum_m \kappa_m g_mY_m Z_m.
\eea
Then, it is easy to show that the $\tilde{\mu}_m$ are invariant under
the transformation $\overline{\mu}_m \rightarrow \overline{\mu}_m +
\alpha \Xi_m$ for an arbitrary constant $\alpha$. By use of
$\tilde{\mu}_m$, $\{n_m\}$ is given by
\beq
   n_m (t_D) =
     \frac{ \kappa_m g_m}{6} \tilde{\mu}_m T_D^2.
\eeq
%
Then, the resultant lepton number density at decoupling is given
by\footnote{More precisely, this result applies only for a flat
  direction which does not include squarks, such as $LH_de$
  \cite{tony}. The result for a flat direction which includes squarks
  will be given in publication \cite{TY}. But, while the forms of $C$
  and $D$ become more complicated, the essential result does not
  change.}
\beq
\label{eq:b-l}
n_{L}(t_D) = \sum_m L_m n_m 
           = - \Delta_R \Delta_L \frac{C}{D} 
                     \frac{T_D^2}{6} \frac{\dot{a}_R}{v_a},
\eeq
where we have defined
\bea
  C &=& R^2 L^2 - \lmk R \cdot L \rmk^2, \non \\ 
  D &=& \Delta_L^2 R^2 + \Delta_R^2 L^2 
        - 2 \Delta_L \Delta_R \lmk R \cdot L \rmk.
\eea
From this result, we can easily reconfirm that an interaction which
has $\Delta_R \ne 0$ and $\Delta_L \ne 0$, that is, which violates
both $PQ$ and lepton symmetries simultaneously, is indispensable for
the final generation of a lepton asymmetry.

As a concrete example of such a violating interaction, we consider the
following dimension five operator,
\begin{equation}
\label{eq:dim5}
{\cal L}_{\not{L}} = \frac{2}{v}l\, l\, H_u H_u +\, {\rm h.c.},
\end{equation}
where $v$ is a scale characterizing the interaction and may be
identified with the heavy Majorana mass for the right-handed neutrino
in the context of the see-saw mechanism. Note that the above
interaction breaks both $PQ$ and lepton symmetries by $\Delta_{R}=2
(R_l + R_{H_u})\ne0$ and $\Delta_{L}=2$. The violating rate of this
interaction is given by $\Gamma \sim 0.04 T^3 / v^2$ \cite{sarkar}.
Then, the decoupling temperature is calculated as
\beq
\label{eq:decT}
T_D \sim 5 \times 10^{11} {\rm GeV} 
             \left(\frac{g_*}{200}\right)^{\frac{1}{2}}
                    \left(\frac{v}{10^{14} {\rm GeV}}\right)^2,
\eeq
where $g_*$ counts the effective degrees of freedom for relativistic
particles.

A part of the produced lepton asymmetry is converted into baryon
asymmetry through sphaleron effects \cite{KSHT}, which is estimated as
\bea
  \frac{n_B}{s} &=& \frac{120 k {\cal R} 
                     ( R_L + R_{H_u})}{23 \pi^2 g_*} 
                      \frac{C}{D}\frac{m_{3/2}}{T_D}\non\\
                &\sim& 3 \times 10^{-10} 
                         \lmk \frac{m_{3/2}}{3 {\rm TeV}} \rmk
                         \lmk \frac{T_D}{10^{12} {\rm GeV}} \rmk^{-1},
\eea
where we have assumed that $\dot{a}_R > 0$ and that $PQ$ charges are
of order unity.

Finally we discuss the constraint on the reheating temperature due to
the gravitino problem. For $m_{3/2} = 3 \sim 10$ TeV, the reheating
temperature is constrained as $T_{RH} \lesssim 10^{12}$ GeV, assuming
that the mass of the lightest supersymmetric particle (LSP) is $O(100$
GeV) \cite{gra}. If we take this bound seriously, the decoupling
temperature $T_D$ must be less than $10^{12}$ GeV so that the
violating interaction is in thermal equilibrium. For $m_{3/2} \sim
100$ GeV, the reheating temperature must be smaller than $T_{RH}
\lesssim 10^{9}$ GeV. However, these constraints on the reheating
temperature are evaded by the introduction of a supersymmetric partner
with a mass much lighter than $100$ GeV. One such particle is the
axino, which naturally exists in our scenario. In fact, it was shown
that the reheating temperature is constrained rather loosely as
$T_{RH} < 10^{15}$ GeV for $m_{3/2} \simeq 100$ GeV, if the axino is
the LSP and the gravitino is the next-to-lightest supersymmetric
particle (NLSP) \cite{asaka}.


In summary, we have discussed a spontaneous baryogenesis mechanism
with an interaction violating another global symmetry in addition to
baryon (lepton) symmetry. We have shown that even if a scalar field
derivatively couples not to baryon current but to another current
associated with some global symmetry $U(1)_Q$, baryogenesis is still
possible by virtue of such a violating interaction. As a concrete
realization of our idea, we have discussed baryogenesis in the context
of a flat direction with neither baryon nor lepton charge.  First, we
have shown that the phase of such a flat direction, strictly speaking,
the NG boson associated with breaking of the symmetry, derivatively
couples to the global current if it is charged under another global
$U(1)_Q$ symmetry, such as the $PQ$ symmetry. The A-term gives this NG
boson a nonzero velocity, which leads to the $CPT$ violation.  By
virtue of the interaction violating $U(1)_Q$ and baryon (lepton)
symmetries simultaneously, $CPT$ violation is transmitted to the
baryon (lepton) sector, which makes it possible to generate baryon
(lepton) asymmetry for this flat direction. In the case of a lepton
asymmetry production, a part of the lepton asymmetry is converted into
the baryon asymmetry through sphaleron effects.

We are grateful to W. Kelly for useful comments and correcting the
English. F.T. is grateful to S. Kasuya and M. Kawasaki for useful
discussions. We are partially supported by the JSPS Grant-in-Aid for
Scientific Research, and M.Y. is partially supported by Department of
Energy under Grant \# DEFG0291ER40688.


\begin{thebibliography}{9}

\bib{Sakharov}
A. D. Sakharov, 
\ZETFP{5}{32}{67}
[\JL{5}{24}{67}].

\bib{CK}
A. G. Cohen and D. B. Kaplan,
\PLB{199}{251}{87}.

\bib{DF}
A. Dolgov and K. Freese,
\PRD{51}{2693}{95};
A. Dolgov, K. Freese, R. Ranagarajan, and M. Srednicki,
\PRD{56}{6155}{97}.

\bib{LFZ}
M. Z. Li, X. L. Wang, B. Feng, and X. M. Zhang,
\PRDD{65}{103511}{02};
M. Li and X. Zhang,
\PLBB{573}{20}{03}.

\bib{FNT}
A. De Felice, S. Nasri, and M. Trodden,
\PRDD{67}{043509}{03}.

\bib{Yamaguchi}
M. Yamaguchi,
\PRDD{68}{063507}{03};
R. H. Brandenberger and M. Yamaguchi,
\PRDD{68}{023505}{03}.

\bib{AD}
I. Affleck and M. Dine,
\NPB{249}{361}{85}.

\bib{Kusenko}
A. Kusenko and M. Shaposhnikov,
\PLB{418}{46}{98}.

\bibitem{Enqvist}
K. Enqvist and J. McDonald,
\NPB{538}{321}{99}.

\bibitem{Kasuya1}
S. Kasuya and M. Kawasaki,
\PRDD{61}{041301}{00}.

\bib{TY}
F. Takahashi and M. Yamaguchi,
hep-ph/0308173.

\bibitem{pq}
R. D. Peccei and H. R. Quinn, 
\PRL{38}{1440}{77}.

\bibitem{DFSZ}
M. Dine, W. Fischler, and M. Srednicki,
\PLB{104}{199}{81};
A. R. Zhitnitsky, 
\SJNP{31}{260}{80}.

\bibitem{Ma}
E. Ma,
\PLBB{514}{330}{01}.

\bibitem{tony}
T. Gherghetta, C. Kolda, and S. P. Martin,
\NPB{468}{37}{96}.

\bibitem{sarkar}
U. Sarkar, 
hep-ph/9809209;
W. Buchmuller, 
hep-ph/0101102.

\bibitem{KSHT}
S. Yu. Khlebnikov and M. F. Shaposhnikov, 
Nucl. Phys. {\bf B308}, 885 (1988);
J. A. Harvey and M. S. Turner, 
Phys. Rev. D {\bf 42}, 3344 (1990).

\bib{gra} 
M. Yu. Khlopov and A. D. Linde, 
\PLBold{138}{265}{84}; 
J. Ellis, E. Kim, and D. V. Nanopoulos, 
{\it ibid}, {\bf 145B}, 181(1984);
M. Kawasaki and T. Moroi, 
\PTP{93}{879}{95};
E. Holtmann, M. Kawasaki, K. Kohri, and T. Moroi,
\PRD{60}{023506}{99}.

\bibitem{asaka}
T. Asaka and T. Yanagida, 
\PLBB{494}{297}{00}.

\end{thebibliography}
\end{document}